\documentclass[%
 ams,
prl,%
 amsmath,amssymb,
 reprint,%
]{revtex4-1}

\usepackage{graphicx}
\usepackage{dcolumn}
\usepackage{bm}

\begin{document}
\begin{abstract}
Superconducting Cu$_x$Bi$_2$Se$_3$ has attracted significant attention as a candidate topological superconductor. Besides inducing superconductivity, the introduction of Cu atoms to this material has also been observed to produce a number of unusual features in DC transport and magnetic susceptibility measurements. To clarify the effect of Cu doping, we have performed a systematic optical spectroscopic study of the electronic structure of Cu$_x$Bi$_2$Se$_3$ as a function of Cu doping. Our measurements reveal an increase in the conduction band effective mass, while both the free carrier density and lifetime remain relatively constant for Cu content greater than $x=0.15$. The increased mass naturally explains trends in the superfluid density and residual resistivity as well as hints at the complex nature of Cu doping in Bi$_2$Se$_3$.
\end{abstract}

\title{Doping-dependent charge dynamics in Cu$_x$Bi$_2$Se$_3$}

\author{Luke J. Sandilands}
\author{Anjan A. Reijnders}
\affiliation{Department of Physics, 60 St. George St., Toronto, Ontario, M5S 1A7, Canada\\}%

\author{Markus Kriener}

\altaffiliation{Current Affiliation: RIKEN Center for Emergent Matter Science (CEMS), Wako 351-0198, Japan\\}

\author{Kouji Segawa}
\author{Satoshi Sasaki}
\author{Yoichi Ando}
\affiliation{Institute of Scientific and Industrial Research, Osaka University, Osaka 567-0047, Japan\\}%

\author{Kenneth S. Burch}
\affiliation{Department of Physics, Boston College, Chestnut Hill, Massachusetts 02467, USA\\}%

\maketitle

The concept of topology in condensed matter physics has attracted renewed interest in recent years due to the discovery of topological insulators, materials where the non-trivial topology of the electronic band structure leads to conducting surface states\cite{thouless1998topological,doi:10.7566/JPSJ.82.102001, RevModPhys.83.1057,hasan2010colloquium}. The principle of band topology soon led to the realization that superconductors can similarly be classified according to the topology of the Bogoliubov-De Gennes Hamiltonian and that topological superconductors could manifest isolated Majorana fermions\cite{RevModPhys.83.1057}. In this context, the discovery of superconductivity in Cu intercalated Bi$_2$Se$_3$, a prototype topological insulator, was notable and this material has attracted significant experimental and theoretical effort \cite{PhysRevLett.104.057001,PhysRevLett.105.097001,PhysRevLett.106.127004,PhysRevLett.109.187003}. ARPES measurements have demonstrated the persistance of the topologically-protected surface states in superconducting Cu$_x$Bi$_2$Se$_3$\cite{Wray:2010uq,PhysRevB.85.125111}, while tunneling and thermodynamic measurements have hinted at an unconventional superconducting state\cite{PhysRevLett.107.217001,PhysRevB.86.180505,1742-6596-449-1-012033,PhysRevLett.108.057001,PhysRevLett.108.057001}. Doping dependent studies revealed an unusual increase in the residual resistivity $\rho_o$ and a suppression of the superfluid density $\rho_s$. The increase in $\rho_o$ was interpreted as an increase in disorder and these two findings therefore suggest unconventional superconductivity in Cu$_x$Bi$_2$Se$_3$\cite{PhysRevB.86.180505}. 

In contrast, the superconducting critical temperature $T_c$ depends only weakly on doping and the carrier density $n$ is completely independent of Cu content at high dopings\cite{PhysRevB.86.180505,PhysRevB.85.125111}. These findings hint at the complex effects of introducing Cu to Bi$_2$Se$_3$. However, the detailed consequences of Cu doping on the electronic structure and disorder levels remains poorly understood. Indeed, a combined ARPES and quantum oscillation study has revealed an open Fermi surface at higher carrier densities\cite{PhysRevB.88.195107}.

To understand the effect of Cu doping on the electronic structure, we have measured the optical response of Cu$_x$Bi$_2$Se$_3$ as function of Cu content $x$. Optical spectroscopy is a well-established tool for investigating the electronic structure of solids\cite{RevModPhys.83.471,dressel2002electrodynamics}. Indeed, optical techniques have been widely applied to topological insulator materials\cite{PhysRevB.89.075138,PhysRevB.86.045439,schafgans2012landau,PhysRevLett.107.136803}, and to Bi$_2$Se$_3$ in particular\cite{PhysRevB.81.125120,Greenaway19651585,PhysRevB.19.1126,PhysRevB.88.075121,PSSB:PSSB2220630116,PhysRevB.87.201201,Wu:2013aa}. However, a systematic study of intra- and interband excitations, as well as disorder levels, as a function of Cu content has not been reported. With this in mind, we have performed reflectance and ellipsometric measurements on a series of Cu$_x$Bi$_2$Se$_3$ samples (x=0.15, 0.22, 0.32, and 0.42). We observe that the plasma minimum, the spectroscopic feature of a metal in reflectance, progressively red-shifts with Cu doping. This result indicates a reduction in the carrier $n$ and/or enhancement in the effective mass ($m_b$) of the carriers with Cu doping. However, the energy of interband transitions from the valence band to the Fermi level ($\epsilon_F$) does not show a concurrent shift. Taken together these results indicate the carrier density $n$ is indeed nearly insensitive to Cu doping for $x\geq0.15$, while the effective mass $m_b$ of the carriers is increased. This result explains the unusual doping dependent evolution of the superfluid density $\rho_s$ and emphasizes the non-trivial effect of the Cu doping.

\section{I.Experimental Details}

Single crystal Cu$_x$Bi$_2$Se$_3$ samples were prepared via the electrochemical intercalation technique described in reference \cite{PhysRevB.84.054513} with x = 0.15, 0.22, 0.32, and 0.42. The superconducting properties were investigated through magnetization measurements. The superconducting critical temperatures (T$_c$) were found to be 3.6, 3.5, 3.3, and 3.1~K respectively, while the shielding fraction ranged from roughly 20 to 40$\%$, consistent with previous reports\cite{PhysRevB.84.054513,PhysRevB.86.180505,PhysRevLett.106.127004}. The Cu content was determined by carefully weighing the samples before and after intercalation.

We carried out reflectance and ellipsometric measurements at 295~K in order to obtain the optical constants of Cu$_x$Bi$_2$Se$_3$. Reflectance measurements were performed at near-normal incidence using a modified Bruker 80v FTIR spectrometer with a series of sources and detectors, as detailed in reference \cite{PhysRevB.89.075138}. Prior to measurement, each sample was freshly cleaved using adhesive tape, meaning our measurements probe charge dynamics in the $ab$-plane. In the case of the x=0.42 sample, the poor sample morphology made cleaving difficult. We therefore prepared a fresh sample surface using a microtome. For x=0.15, 0.32 and 0.42, we measured the reflectance from 100 meV $\rightarrow$ 805 meV. This range was extended to 5 meV $\rightarrow$ 1.24 eV in the case of the x=0.22 sample. We used an $in$ $situ$ gold overcoating technique to provide an absolute intensity reference for our measurements\cite{Homes:93}. The dielectric function $\hat\epsilon(\omega)$ of all four samples was then measured directly for the range 0.75 eV to 5 eV using a Woolam VASE ellipsometer. Importantly, the reflectance computed from the ellipsometric data is in good agreement with the measured reflectance in the frequency region where the two data sets overlap. We also checked the quality of the micro-tomed x=0.4 surface by varying the polarization of incident light in our reflectance measurement and the angle of incidence in ellipsometry. Neither variation produced a meaningful change in the data, validating our surface preparation technique. 

To extract  the dielectric function $\hat\epsilon(\omega)$ from the optical data, we employ a variational dielectric function implemented in the Reffit software package\cite{kuzmenko:083108,reffit}. We simultaneously fit the reflectance and ellipsometric data in order to constrain the phase of $\hat\epsilon(\omega)$at high energies\cite{PhysRevB.42.1969,PhysRevB.89.075138}. In considering the optical response of our sample, we focus on the imaginary part of the dielectric function $\epsilon_2(\omega)$, which represents charge excitations in a solid\cite{yu2010fundamentals}. Finally, to make connections between the optical response and transport measurements, $\epsilon_2(\omega)$ can also be recast in terms of the real part of the optical conductivity $\sigma_1(\omega) = \omega\epsilon_2(\omega)/4\pi$~(CGS units), the finite frequency generalization of the DC conductivity.

\section{II. Broad band optical response}

Cu$_x$Bi$_2$Se$_3$ exhibits the optical response characteristic of a heavily-doped metallic semiconductor. Before discussing the data in detail it is therefore worthwhile to review the semiclassical Drude model, the common starting point for understanding the charge dynamics of a metal\cite{yu2010fundamentals,dressel2002electrodynamics}. Within the Drude model, the dielectric function $\hat{\epsilon} = \epsilon_1 +i\epsilon_2$ is given by:

\begin{equation}\label{drude}
\hat{\epsilon}(\omega) = \epsilon_{\infty} - \frac{\omega_p^2}{\omega^2+i\omega(1/\tau)}.
\end{equation}

Here $1/\tau$ is the free carrier scattering rate, $\epsilon_{\infty}$ represents screening due to interband transitions located at higher energies, and  $\omega_p$ is the plasma frequency and is proportional to $(n/m_b)^{1/2}$, where $n$ is the carrier density and $m_b$ is the optical band mass. A Drude analysis therefore allows access to parameters characterizing both the electronic structure ($\omega_p$, $\epsilon_\infty$) and the disorder level ($1/\tau$). 

It is also informative to consider the low and high frequency limits of this model dielectric function. At low frequencies, $\epsilon_1$ is large and negative. Accordingly, the reflectance in a metal is high at low frequencies. For $\omega_p \gg 1/\tau$, the real part of the dielectric function $\epsilon_1$ has a zero crossing at the screened plasma frequency $\widetilde{\omega}_p = \omega_p/\sqrt{\epsilon_{\infty}}$. This leads to a characteristic dip in the reflectance known as the plasma minimum whose line shape is determined by $1/\tau$. The location and width of the plasma minimum therefore provide insight into the free carrier dynamics. 

\begin{figure}
 \includegraphics[width=\columnwidth]{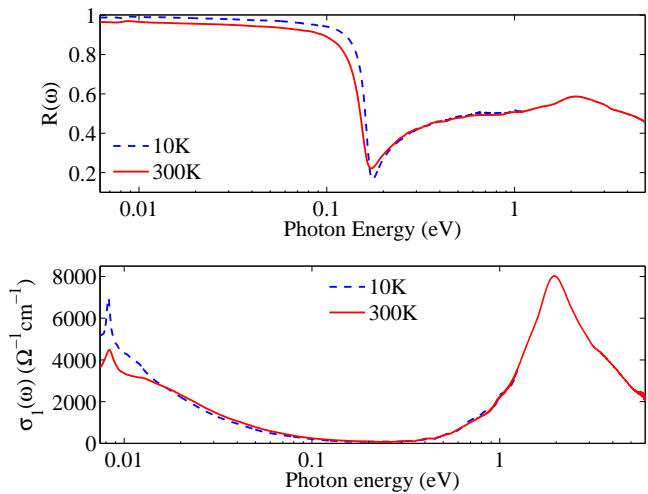}%
 \caption{\label{fig:1}  Optical properties of Cu$_{0.22}$Bi$_2$Se$_3$ at 10 and 300 K. Top panel: Reflectance. Bottom panel: Optical conductivity $\sigma_1 (\omega)$. Both spectra are compatible with a heavily-doped, metallic semiconductor.}%
 \end{figure}

In the top panel of figure \ref{fig:1} we display the broadband optical response of Cu$_{0.22}$Bi$_2$Se$_3$ at 295K. Several features merit mention. At low frequencies, the reflectance is dominated by a Drude line shape with a well-defined plasma minimum near 0.2 eV. This is in rough agreement with previous optical studies of Cu$_x$Bi$_2$Se$_3$ that reported plasma minima between 70 and 200 meV depending on Cu content\cite{PhysRevB.87.201201,PhysRevB.19.1126}. The slight kink below 0.01 eV is the $\alpha$ phonon typically observed in Bi$_2$Se$_3$\cite{PhysRevB.81.125120}, which we do not discuss further. The slope of the reflectance also shows a weak maximum near 2 eV due to interband transitions\cite{PhysRevB.81.125120}.

These various features are more readily discerned from the optical conductivity $\sigma_1(\omega)$ shown in the bottom panel of figure \ref{fig:1}. A strong, narrow peak due to the $\alpha$ phonon can be seen on top of a broad Drude contribution due to free carriers. At 0.35 eV, $\sigma_1$ rises sharply as excitations from the valence band to the conduction band become energetically allowed. The low energy dynamics are well-described by a single Drude lineshape at both 10 and 300~K and therefore with a single species of carriers. In contrast, a previous optical study of Cu$_{0.07}$Bi$_2$Se$_3$\cite{PhysRevB.87.201201} identified a peak at finite (25 meV) energy, indicating the presence of a bound impurity states, as well as a Drude component. Typically such impurity-related features merge with the Drude component as the free carrier density increases\cite{PhysRevLett.71.3681,PhysRevB.52.16486}. Since the carrier density in our crystals is an  order of magnitude higher than reported in reference \cite{PhysRevB.87.201201}, the absence of the impurity-related feature in our data is not surprising. A second possibility is that the lack of an impurity peak in our data is simply due to differences in crystal preparation. Furthermore, Dordevic et al.\cite{0953-8984-25-7-075501} interpreted the free carrier contribution to $\hat{\epsilon}(\omega)$ in a variety of topological insulator materials in terms of carrier density inhomogeneity. Our observation of a single Drude mode, even at 10~K, suggests that such inhomogeneities are small in our samples, at least in the x = 0.22 and at the temperatures and frequencies measured. The interband transitions are similar to those observed in pristine Bi$_2$Se$_3$, albeit with a Moss-Burstein shifted direct gap\cite{PhysRevB.81.125120,PhysRev.93.632,0370-1301-67-10-306}. 

 \makeatletter 

\section{III. Doping dependence of the plasma minimum and interband transitions}

\begin{figure}
 \includegraphics[width=\columnwidth]{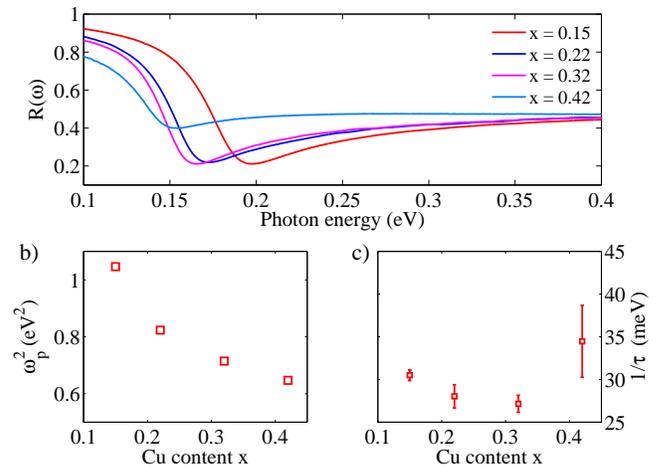}%
 \caption{\label{fig:2}  Doping dependence of the free carrier dynamics in Cu$_x$Bi$_2$Se$_3$. a) Mid-infrared reflectance b) Squared plasma frequency $\omega_p^2$. c) Scattering rate $1/\tau$. The error bars in $\omega_p$ are smaller than the symbols used in b). The reflectance minimum $\widetilde{\omega}_p$ is seen in panel a) to red-shift with doping, suggesting a change in carrier dynamics. This is borne out by the fitted values of $\omega_p^2$, which red-shifts by roughly 40$\%$. In contrast, $1/\tau$ shows only a slight increase at high doping. The change in $\omega_p$ is due to an increase in the band mass $m_b$.}%
 \end{figure}

In figure \ref{fig:2}(a), we show the mid-infrared $ab$-plane reflectance for samples with different doping levels. The location of the plasma minimum ($\widetilde{\omega}_p$) red-shifts with doping and implies a change of the in-plane carrier dynamics. To quantitatively study the evolution of the carrier dynamics, we fit our reflectance data near the plasma minimum (111 meV to 600 meV) to a Drude model.  
In figure \ref{fig:2}(b) and (c) we show the $\omega_p^2\ (\propto n/m_b)$ and $1/\tau$ values derived from this procedure. As expected from the reflectance data, $\omega_p^2$ red-shifts by $40\%$ over the measured doping range, while $1/\tau$ is relatively constant before sharply increasing by $20 \%$ for the x=0.4 sample. For comparison, we note that previous optical spectroscopy measurements of pristine Bi$_2$Se$_3$ found a room temperature $1/\tau$ ranging from $1$ to $8$ meV, significantly lower than our values\cite{PhysRevB.81.125120,PhysRevB.81.241301}. These works also reported $\omega_p$ values between $47$ and $197$ meV, also  lower than our values which range from $0.6$ to $1$ eV. This indicates that both $n$ and disorder are increased relative to the pristine compound. Returning to \ref{fig:2}(b), we note that $\omega_p$ decreases with increased Cu content.  The shift of $\omega_p$ with doping signifies a change in the quantity $n/m_b$, where $m_b$ is the $ab$-plane optical band mass. Hall effect\cite{PhysRevB.86.180505} and ARPES\cite{PhysRevB.85.125111} measurements have shown that $n$ is effectively constant over the measured doping range, and so our data indicates an increase in $m_b$ with Cu doping.

 \makeatletter 

 \makeatletter 
 \begin{figure}
 \includegraphics[width=\columnwidth]{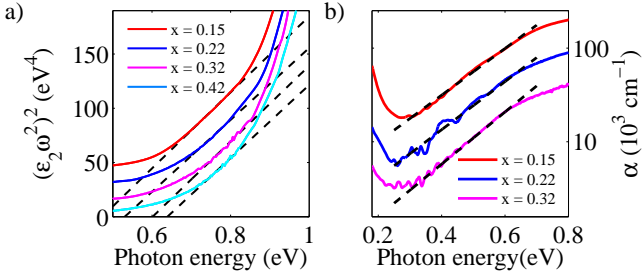}%
 \caption{\label{fig:3} Interband transitions in Cu$_{x}$Bi$_2$Se$_3$. a) $(\epsilon_2\omega^2)^2$ v.s photon energy. The 0.15, 0.22, 0.32 curves have been offset by 45, 30 and 15 eV$^4$ respectively. b) Absorption $\alpha(\omega)$ vs. photon energy. The 0.15 and 0.22 curves have been multiplied by factors of 4 and 2 for clarity. The linear behavior in $(\epsilon_2\omega^2)^2$  indicates direct optical transitions. Below this energy, $\alpha(\omega)$ follows an exponential (Urbach) form due to disorder.}
 \end{figure}

 \makeatletter 
 \begin{figure}
 \includegraphics[width=\columnwidth]{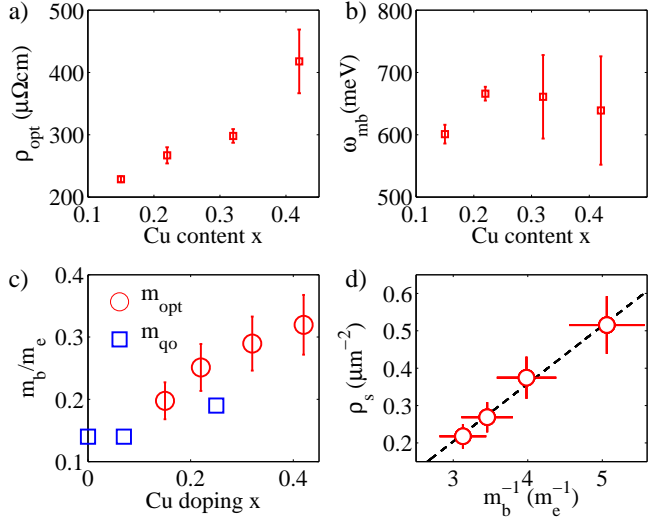}%
 \caption{\label{fig:4} Doping dependent charge dynamics in Cu$_x$Bi$_2$Se$_3$. a) Optical resistivity $\rho_{opt}$.  b) Moss-Burstein edge $\omega_{mb}$. c) Band mass $m_{b}$ from optical ($m_{opt}$) and quantum oscillation ($m_{qo}$) measurements. d) Superfluid density $\rho_s = \lambda^{-2}$ versus inverse bandmass $m_b^{-1}$. $\rho_{opt}$ shows an increase with doping, consistent with the behavior of the residual resistivity observed in DC transport\cite{PhysRevB.86.180505}. $\omega_{mb}$, and therefore $n$, does not vary systematically with doping, while $m_b$ can be seen to increase. $\rho_s$ and $1/m_b$ show a linear relationship, indicating that the doping dependence of $\rho_s$ is due to the increased $m_b$, rather than disorder. }%
 \end{figure}

In order to confirm a constant $n$ in our samples, we now consider the interband transitions present in Cu$_x$Bi$_2$Se$_3$. In a doped semiconductor, the onset of interband transitions is increased from the bare band gap $E_g$ by the Fermi level $\epsilon_F$, an effect known as the Moss-Burstein shift\cite{PhysRev.93.632,0370-1301-67-10-306}. In the simplifying case of a direct gap semiconductor with parabolic bands, the Moss-Burstein edge $\omega_{mb}$ is expected to vary as $\omega_{mb} = E_g + (1+m_c/m_v)(\epsilon_F-4k_bT)$, where $E_g$ is the optical gap of pristine Bi$_2$Se$_3$, $m_c (m_v)$ is the conduction (valence) band mass, and $\epsilon_F$ is the Fermi level measured from the bottom of the conduction band\cite{PhysRev.93.632}. The onset of interband transitions $\omega_{mb}$ therefore provides a measure of $\epsilon_F$. We note that this picture of a relatively rigid band structure is consistent with a recent ARPES study which found that the principal effect of Cu doping is to shift $\epsilon_F$\cite{PhysRevB.85.125111}.

As can be seen in the bottom panel of figure \ref{fig:1}, interband transitions become important above 0.35 eV. For x=0.15-0.32, these features are consistent with 0.6 eV excitations from the valence band to $\epsilon_F$ that are broadened by disorder. In figure \ref{fig:3}(a), we plot the quantity $(\epsilon_2 \omega^2)^2$ which shows a linear variation in energy between near 0.8 eV. This is a signature of the onset of direct optical transitions between parabolic bands and the intercept with the energy axis gives the value of the gap\cite{yu2010fundamentals}. The gap values  suggested by this analysis are shown in figure \ref{fig:4}(b). The large blue-shift of the direct gap with respect to the 115-150 meV optical gap of pristine Bi$_2$Se$_3$\cite{PSSB:PSSB2220630116,Greenaway19651585} at 295K is due to the the Moss-Burnstein effect\cite{0370-1301-67-10-306,PhysRev.93.632}. This interpretation is also consistent with photoemission. Specifically, ARPES suggest that the lowest energy direct optical transition from the valence band to $\epsilon_F$ should occur at roughly 0.6-0.7 eV\cite{PhysRevB.85.125111,Wray:2010uq}. 

Interband transitions are also sensitive to disorder. Indeed, below below the gap (roughly 0.6 eV), the absorption $\alpha(\omega)$ in Cu$_x$Bi$_2$Se$_3$ follows an exponential form as can be seen in figure \ref{fig:3}(b). This in-gap 'Urbach tail' is a well-known phenomenon in semiconductors and is caused by structural and thermal disorder\cite{PhysRev.92.1324,PhysRevB.51.1778}. The absorption in this regime is typically parameterized as ln($\alpha(\omega)) \propto -(\omega-\omega_{mb})/\sigma_o$ where $\omega_{mb}$ is energy of the transition and  $\sigma_o$ is the Urbach parameter that determines the width of the exponential region and is related to the disorder. The x=0.15, 0.22, and 0.32 samples show Urbach tail behavior.  For x=0.42, we observe significant in-gap weight that does not follow the Urbach tail form. 

\section{IV. Discussion}

In table \ref{params} we summarize the parameters characterizing the electronic structure of Cu$_x$Bi$_2$Se$_3$ for x=0.15-0.42. We have also included x=0 parameters from reference \cite{PhysRevB.81.125120}. $\omega_{mb}$ and $\sigma_o$ are obtained from fits to the direct gap and Urbach expressions described previously. These parameters reveal two doping regimes. At low doping (below x=0.15), the carrier density (in terms of $\omega_p$ and $\omega_{mb}$) and the disorder (in terms of $1/\tau$) rise rapidly with respect to the pristine compound. Above this level, $\omega_{mb}$ and $1/\tau$ saturate, while $\omega_p$ decreases by 20 $\%$. The saturation of $\omega_{mb}$ suggests that $n$ does not vary systematically for x $>$ 0.15 and is relatively constant, consistent with Hall\cite{PhysRevB.86.180505} and ARPES\cite{PhysRevB.85.125111} measurements in similar crystals. Indeed, taking $E_g =0.15$ eV\cite{Greenaway19651585}  and $ m_c/m_v =1$\cite{PSSB:PSSB2220630116} yields an $\epsilon_F$ of 326-356 meV for our samples, in good agreement with photoemission experiments which suggest values between 290 and 350 meV\cite{PhysRevB.85.125111,PhysRevB.88.195107}. With reference to figure \ref{fig:3}b), we also observe that the maximum allowed decrease in $\epsilon_F$ is $\delta \epsilon_f =$ 32 meV. Using this fact, we can estimate the largest possible decrease in $n$ allowed by our data. Given that $\epsilon_f \propto  n^{2/3}$, $\delta n/n \approx 3\delta \epsilon_f/2\epsilon_F = $3(30 meV)/2(326 meV) $= 0.15$

The observed behaviour of $\epsilon_F$ and $n$ is in stark contrast to the decrease of $\omega_p$ with doping above x=0.15. In particular, a $15\%$ decrease in $n$ is insufficient to explain the observed decrease in $\omega_p^2$ ($\propto n/m_b$) of $40\%$. Our results therefore demonstrate a change in the band mass $m_b$ with increased Cu content, rather than a simple change in $n$. In figure \ref{fig:4}(c), we show the $m_b$ implied by our data. We assumed a free carrier density of $1.5 \pm 0.4 \times 10^{20}$ cm$^{-3}$ after reference \cite{PhysRevB.86.180505}, which results in the relatively large error bars. Nonetheless, our results are in good agreement with the $m_b$ values (obtained from quantum oscillation measurements) reported in the literature, which are included for comparison in figure \ref{fig:4}(c) \cite{PhysRevB.81.195309, PhysRevB.87.201201, PhysRevLett.109.226406}. 

Our measurements show that the band mass progressively increases with Cu content, reaching a value roughly twice that of pristine Bi$_2$Se$_3$ at x=0.42. We also note that Lahoud et al.\cite{PhysRevB.88.195107} reported quantum oscillation measurements suggesting a $m_{b} = 0.25m_e$ for heavily-doped Cu$_x$Bi$_2$Se$_3$, although they do not specify the Cu content $x$. These authors also reported ARPES data indicating that the in-plane band curvature at $\epsilon_F$ is reduced (consistent with an increased mass) as $k_z$ moves away from the Gamma point in heavily Cu doped samples. Finally, an $m_b$ of $0.25m_e$ has also been identified though previous optical measurements in heavily-doped Cu$_x$Bi$_2$Se$_3$ and interpreted in terms of band non-parabolicity\cite{PhysRevB.19.1126}. While this value is in good agreement with our results, our study has revealed that mass enhancement can occur without significant change in $n$. 

The variation in $m_b$ that we observe naturally explains the unusual doping dependence of the residual resistivity $\rho_o$ measured by DC transport\cite{PhysRevB.86.180505}.  In figure \ref{fig:4}(a), we show the DC resistivity $\rho_{opt}$ implied by our data. This quantity increases by roughly a factor of two over the measured doping range, similar to $\rho_o$\cite{PhysRevB.86.180505}. Given that $m_b$ increases by a comparable factor while $1/\tau$ is relatively unchanged above $x=0.15$, we can attribute the increased $\rho_{opt}$ and $\rho_{o}$ to the increased $m_b$, rather than an increase in disorder. 

The change in band mass also accounts for the anomalous evolution of the superfluid density with doping without invoking disorder and unconventional pairing\cite{PhysRevB.86.180505}. The superfluid density $\rho_s$ can be defined as 1/$\lambda^2 \propto m_b^{-1}$, where $\lambda$ is the superconducting penetration depth\cite{tinkham}. In \ref{fig:4}(d), we plot $\rho_s$ (reproduced from Kriener et al.\cite{PhysRevB.86.180505}) versus our measured $m_b^{-1}$, demonstrating a clear linear relationship. Heuristically, the Cooper pairs are becoming heavier and so the condensate is less effective in screening an applied magnetic field leading to an increased $\lambda$. More generally, this trend is a manifestation of the Homes' superfluid scaling relation, which states that $\rho_s \propto \sigma_1(0)T_c$\ \cite{Dordevic:2013fk}. Since $\sigma_1(0) \propto m_b^{-1}$, an increase in $m_b$ reduces the low energy spectral weight available to the condensate and so decreases $\rho_s$. We also note that Cu$_x$Bi$_2$Se$_3$ is firmly in the dirty limit in the doping range studied, with $1/\tau \sim 30$ meV much larger than the superconducting gap $\Delta \sim 0.6$ meV estimated from $T_c$\cite{Wray:2010uq}.  

The conclusions reached above regarding the low temperature properties of Cu$_x$Bi$_2$Se$_3$ are based primarily on data collected at 295~K. However, we believe that this is justified for several reasons. First, the low temperature optical data shown in figure \ref{fig:1} do not evince a significant change in $\omega_p$, suggesting the 295~K data are a good reflection of the low temperature dynamics. Second, as can be seen in \ref{fig:4}(c), our measured values of $m_b$ are in good agreement with quantum oscillation results obtained at low temperatures and energies for crystals of similar dopings. The Hall effect and resistivity data for a number of doping levels reported in reference \cite{PhysRevB.84.054513} also confirm that the low energy electronic structure is not significantly modified at low temperatures. Moreover, $\rho(4\ K)/\rho(295\ K)$ was found to be $\sim 0.5$ and so we do not expect $1/\tau$ to change by more than a factor of 2 at low temperature.

The origin of the increased $m_b$ is not clear. One possibility is that Cu intercalation distorts the host Bi$_2$Se$_3$ lattice and reduces the effective in-plane bandwidth. Indeed, Cu intercalation is known to increase the $c$-axis lattice constant\cite{PhysRevLett.104.057001,PhysRevB.87.201201} and so would naively be expected to alter the electronic dispersion in this direction. Any effective in-plane hoppings which involve intermediate steps between quintuple layers would therefore be similarly reduced, leading to an increased in-plane mass. 

The mass enhancement could also be a many-body effect, specifically band gap renormalization\cite{0022-3719-9-7-009,PhysRevB.24.1971,PhysRevB.66.201403}. Accounting for many body effects, the zero temperature measured optical band gap $\omega_{opt}$ of a doped semiconductor is given by  $\omega_{opt} = E_g -\Delta_{RN}+(1+m_c/m_v)\epsilon_F$, where $\Delta_{RN}$ is the band gap reduction due to many-body effects\cite{1367-2630-15-7-075020} and $(1+m_c/m_v)\epsilon_F$ is the Moss-Burstein shift. This band-gap reduction would also be expected to influence the conduction band mass, such as predicted by $k\cdot p$ theory\cite{yu2010fundamentals}. A possible explanation of our data is therefore that the band gap renormalization nearly cancels the Moss-Burstein shift, resulting in an almost constant optical band gap, and also leads to an increased effective mass. 

Our results also point to the complex nature of Cu doping. Previous work in this direction has emphasized the ambipolar nature of the Cu dopant\cite{PhysRevLett.104.057001,vasko_amphoteric}. In particular, Cu atoms are thought to enter the Bi$_2$Se$_3$ lattice in two principle ways: in the van der Waals gap, where each dopant atom donates a single electron, and substitutionally on a Bi site, where the dopant instead decreases the free electron concentration\cite{vasko_amphoteric}. In this context, one would naively expect two possibilities upon progressive Cu doping. If the dopant atoms continue to enter the van der Waals gap, then $n$ should increase monotonically, in contrast to experiment. On the other hand, if at high concentrations some fraction of the dopants begin occupying Bi sites such that $n$ remains constant, then we would expect a drastic increase in the in-plane scattering rate $1/\tau$. However, as can be seen in table \ref{params} both $1/\tau$ and $\sigma_o$, two measures of disorder, are effectively constant above $x = 0.15$. This suggests that Cu atoms do not substitute for Bi in large numbers. Indeed, our optical spectroscopy reveal that beyond simply adding carriers or introducing disorder, Cu doping impacts the electronic dispersion, as evinced by the increased $m_b$. Further work is therefore required to elucidate in detail the effect of Cu doping at high concentrations. For instance, channeling experiments, such as have been performed on (Ga,Mn)As\cite{Dobrowolska:2012fk}, would be helpful in determining the precise location of the Cu atoms within the Bi$_2$Se$_3$ lattice. Scanning tunnelling microscopy and $ab\ initio$ calculations have recently been used to identify a variety of dopant positions in Cu$_x$Bi$_2$Se$_3$\cite{PhysRevB.89.155312} and would also be useful to investigate any local changes in electronic structure associated with these various dopant locations.

Finally, the fact that $m_b$ is only modestly increased by Cu doping up to $m_b \approx 0.3 m_e$ at x =0.4, is in apparent contradiction with the mass $m_{sh} = 2.6 m_e$ inferred from the electronic specific heat\cite{PhysRevLett.106.127004}. However, the specific heat data was interpreted under the assumption that the density of states at $\epsilon_F$ is given by the free electron model. Besides the possibility of many-body effects suggested by our optical studies, recent ARPES experiments have demonstrated that spin-plasmon excitations cause a giant surface state mass enhancement in Cu$_x$Bi$_2$Se$_3$\cite{PhysRevLett.110.217601}. This means that the free electron model is possibly inadequate for describing the density of states at $\epsilon_F$ and so the $m_{sh}$ derived from such a treatment should be interpreted with caution.

\begin{table}\label{params}
\begin{tabular}{ c | c c  c  c  c c}
  \hline
   x   & $\omega_p$ (meV) & $1/\tau$ (meV) & $\omega_{mb}$ (meV) & $\sigma_0$ (meV)  \\
   \hline \hline
   0        & 197  & 8  & 255$\pm$19 &$-$\\
   \hline 
   0.15        & 1023$\pm$5  & 30.5 $\pm$0.6 & 601$\pm 15$ &175$\pm$1\\ 
   \hline
    0.22    & 907$\pm$8  &  28.0$\pm$1.4 & 666$\pm$11 & 169$\pm$1 \\
   \hline
       0.32    & 845$\pm$6  &  27.2$\pm$1.0 & 661$\pm$67 & 159$\pm$1\\
   \hline
      0.42   & 804$\pm$5  &  34.5$\pm$4.2 & 639$\pm$87 & $-$ \\
\end{tabular}
\caption{\label{params} Parameters characterizing the electrodynamics of Cu$_x$Bi$_2$Se$_3$. The parameters are the plasma frequency $\omega_p$, the free carrier scattering rate $1/\tau$, the Moss-Burnstein edge $\omega_{mb}$, and the Urbach parameter $\sigma_o$. Between $x=0$ and $x=0.15$, $\omega_p$, $\omega_{mb}$, and $1\tau$ increase significantly. This implies that both $n$ and the disorder level are increased relative to the pristine compound.  At higher doping levels, however, $\omega_{mb}$ and $1/\tau$ do not vary systematically, within our experimental accuracy, implying a constant $n$ and disorder level. Taken together with the decrease in plasma frequency $\omega_p$, this indicates an increase in the band mass with Cu doping x. The $x=0$ data are reproduced from reference \cite{PhysRevB.81.125120}.}
\end{table}

\section{V. Conclusions}

In summary, we have studied the doping dependent charge dynamics of the candidate topological superconductor Cu$_x$Bi$_2$Se$_3$ using optical spectroscopy. We observe a non-monotonic evolution of $\omega_p$ with doping. While both $n$ and the disorder levels are increased in our samples relative to the pristine compound, our analysis shows that these quantities are relatively constant above x=0.15. Given a constant $n$, the decrease in $\omega_p$ therefore suggests an increase in the optical band mass $m_{b}$ at high dopings. The increased $m_b$ accounts for the apparent suppression of $\rho_s$ and the increase in $\rho_o$ with doping. Our results highlight the non-trivial nature of Cu doping in Bi$_2$Se$_3$.

\begin{acknowledgements}
We would like to acknowledge Peter Brodersen at Surface Interface Ontario at the University of Toronto for assistance with the microtome. The work at the University of Toronto was funded by the Ontario Research Fund, NSERC, and Canada Foundation for Innovation. The work at Osaka was supported by JSPS (KAKENHI 25220708 and 24540320), MEXT (Innovative Area `Topological Quantum Phenomena' KAKENHI), AFOSR (AOARD 124038), and Inamori Foundation. M.K. was supported by JSPS (KAKENHI 25800197).
\end{acknowledgements}

%

 \end{document}